# Enhanced third harmonic generation with graphene metasurfaces

Boyuan Jin, Tianjing Guo, and Christos Argyropoulos*

Department of Electrical and Computer Engineering, University of Nebraska-Lincoln, Lincoln, NE, 68588, USA

*christos.argyropoulos@unl.edu

**Abstract**

The nonlinear responses of different materials provide useful mechanisms for optical switching, low noise amplification, and harmonic frequency generation. However, the nonlinear processes usually have an extremely weak nature and require high input power to be excited. To alleviate this severe limitation, we propose new designs of ultrathin nonlinear metasurfaces composed of patterned graphene micro-ribbons to significantly enhance third harmonic generation (THG) at far-infrared and terahertz (THz) frequencies. The incident wave is tightly confined and significantly boosted along the surface of graphene in these configurations due to the excitation of highly localized plasmons. The bandwidth of the resonant response becomes narrower due to the introduction of a metallic substrate below the graphene micro-ribbons, which leads to zero transmission and standing waves inside the intermediate dielectric spacer layer. The enhancement of the incident field, combined with the large nonlinear conductivity of graphene, can dramatically increase the THG conversion efficiency by several orders of magnitude. In addition, the resonant frequency of the metasurface can be adjusted by dynamically tuning the Fermi energy of graphene via electrical or chemical doping. As a result, the third harmonic generated wave can be optimized and tuned to be emitted at different frequencies without the need to change the nonlinear metasurface geometry. The proposed nonlinear metasurfaces provide a new way to realize compact and efficient nonlinear sources at the far-infrared and THz frequency ranges, as well as new frequency generation and wave mixing devices which are expected to be useful for nonlinear THz spectroscopy and noninvasive THz imaging applications.

## 1. Introduction

The far-infrared and terahertz (THz) frequency regions have wide applications in noninvasive imaging, spectroscopy, wireless communications, sensing, quality control, and security technologies [1-5]. Nonlinear optical effects in these frequency ranges have drawn increased attention, since they offer the possibility to realize several new optical parametric processes, such as generation of waves at new frequencies and ultrafast manipulation of light with light [6]. However, the nonlinear effects suffer from very weak efficiencies, usually requiring enormous input radiation intensities to be exited, making their practical applications challenging if not impossible. The recent research area of plasmonics consists an attractive platform to overcome the intrinsic weak nature of nonlinear effects [7]. In principle, plasmonic devices are able to extremely confine the optical power in subwavelength regions and, thus, create locally confined strong electromagnetic fields. In addition, they also have the advantages of ultrafast response time and high sensitivity to the surrounding dielectric media [7-10].

Plasmonic systems are usually based on metals and operate at high optical frequencies (visible and near-IR). Recently, it was demonstrated that graphene is the ideal platform to achieve plasmonic behavior at lower far-infrared and THz frequencies [11]. It is a monolayer two-dimensional (2D) material with one atom thickness, exhibiting many appealing electrical, optical, mechanical, and thermal properties. Compared to noble metals, graphene has smaller electron density, and consequently lower plasma frequency located at the far-infrared and THz frequency spectrum [12-15]. Furthermore, theoretical and experimental works have demonstrated that the third order nonlinearity of graphene at the THz frequency range is exceptionally strong, and, at the same time, graphene has smaller losses with respect to noble metals [16-21]. These features make graphene a promising candidate to enhance nonlinear effects at the far-infrared and terahertz frequencies. The wavelength of surface plasmons along graphene can be 40-100 times smaller compared to free space, which paves the way to implement highly subwavelength planar on chip THz devices [13, 21-24]. Besides the strong plasmonic resonant response, the optical fields can be further enhanced by patterning graphene forming graphene-based metasurfaces [25-27]. Chemical vapor deposition (CVD) and electron beam lithography can be used to fabricate patterned graphene microstructures [28, 29].

However, the efficiency of several nonlinear parametric processes, including various wave mixing phenomena, is limited by phase-matching conditions. By compressing the volume of light-matter interactions to extremely subwavelength regions, the phase-matching requirements along graphene can be relaxed, as it was shown before by using nonlinear metallic and dielectric metasurfaces [30-33]. Moreover, among the many appealing properties of graphene, one of the most useful properties is that both its linear and nonlinear conductivities are highly tunable via electrical or chemical doping [34-37]. The tunability in their response can make graphene metasurfaces to adapt to different applications after their fabrication without altering their geometry and structure. In this respect, graphene is more advantageous and more flexible to be used at several applications compared to usual plasmonic materials, i.e. noble metals.

In this work, we focus our attention on ways to enhance the third harmonic generation (THG) nonlinear process at far-infrared and THz frequencies. Compared to four wave mixing (FWM), which is another common third-order nonlinear frequency-mixing procedure [6], THG can realize large frequency conversion using only one input source, leading to lower background noise and cubic power dependence of the nonlinear output signal to the input signal [16]. In order to enhance THG, a novel resonant nonlinear metasurface is proposed and designed based on periodic graphene micro-ribbons placed on top of a dielectric layer and terminated by a metallic substrate. Numerical simulations of different nonlinear graphene structures have been previously implemented based on the finite-difference time-domain (FDTD) or the boundary-integral spectral element methods [9, 38, 39]. In this work, we develop a full-wave nonlinear model using COMSOL Multiphysics, a commercial electromagnetic solver based on the finite element method (FEM), to accurately compute the nonlinear response of the proposed graphene metasurface in the frequency domain. FEM is the ideal numerical technique to simulate optical nonlinear effects because it does not suffer from numerical instabilities, which are common problem in time-domain numerical methods

(FDTD). In addition, it uses a tetrahedral mesh that produces more accurate results when curved geometries are needed to be simulated and is faster and less memory intensive compared to other numerical techniques.

Based on the presented numerical analysis, it is revealed that the fields at the input fundamental frequency (FF) resonant mode can be tightly confined and significantly enhanced along the graphene micro-ribbons surface due to the excitation of highly localized THz surface plasmon polaritons (SPP). Besides, standing wave effects based on Fabry-Pérot (FP) resonances are obtained due to the strong reflection from the bottom metallic substrate, which causes further enhancement of the electric field and a sharper resonance response at the FF [27, 40]. As a result, the THG efficiency can be increased by several orders of magnitude when the FF is close to the resonance of the metasurface. We investigate the dependence of the third harmonic (TH) radiation on the FF, incident angle, input intensity, and several other graphene parameters, such as mobility and Fermi energy. In addition to the formation of strong FP resonances, the backside metal substrate can also act as a gate electrode enabling the tunability of graphene's Fermi energy via the applied bias voltage. The resonant frequency of the proposed device can be adjusted by dynamically tuning the Fermi energy or doping level and the nonlinear metasurface can emit TH radiation at different frequencies without changing its geometry and dimensions. The presented results demonstrate that among all the different materials of the proposed metasurface, the nonlinear properties of the patterned graphene micro-ribbons are the dominant contributors to the THG process. Compact and efficient nonlinear sources, new harmonic frequency generators, and wave mixing devices are envisioned based on the proposed nonlinear graphene metasurfaces operating at the far-infrared and THz frequency ranges.

## 2. Geometry and theoretical model

The geometry of the proposed nonlinear graphene metasurface is shown in figure 1. An array of graphene micro-ribbons is placed over a thin dielectric spacer layer that is terminated by a metallic substrate. The spacer layer between graphene micro-ribbons and metal termination is filled with a Kerr nonlinear dielectric material. The linear and nonlinear responses of the metasurface are computed using the electromagnetic solver of COMSOL Multiphysics. Assuming that the length of the graphene micro-ribbons is much larger than their width, the proposed metasurface is uniform along the direction of z-axis and two-dimensional simulations are used to calculate its linear and nonlinear responses. The simulation domain is terminated by periodic boundary conditions normal to the x-axis with an optimized periodicity of $a = 3.88$ μm. Thus, only one graphene ribbon needs to be contained in the simulation domain, with a ribbon width equal to $d = 0.75a$. We picked these geometric values in order to make the metasurface resonate at 3THz, as it will be shown in the next section. The thickness of the backside metal, which is assumed to be gold, is chosen to be much larger than its THz skin depth. Hence, the metallic layer acts as a perfect reflecting mirror and zero transmission is obtained. The resonant condition of the proposed metasurface becomes sharper and enhanced due to the formation of a cavity that introduces FP standing waves inside the dielectric layer, independent of the spacer layer thickness [27, 40]. In addition, the metallic substrate can also serve as the gate electrode to electrically dope graphene leading to tunable Fermi energy via the applied bias voltage.

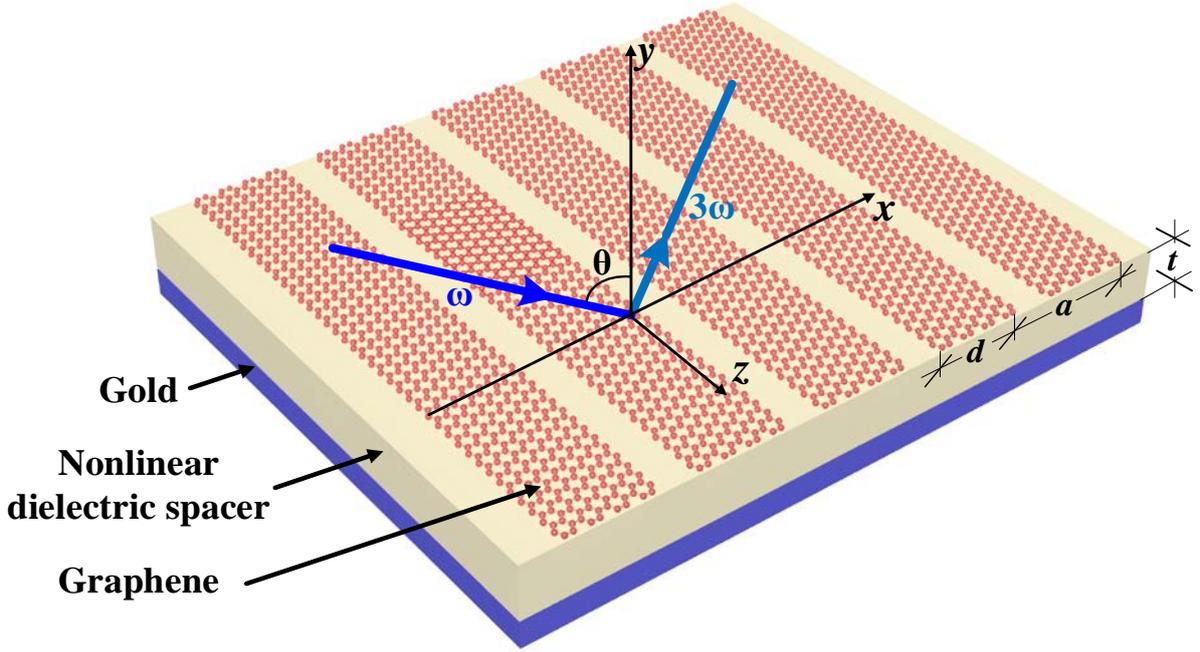

Figure 1. Schematic illustration of the proposed nonlinear metasurface.

During the THG process, three photons with the same fundamental angular frequencies $\omega_{FF}$ are annihilated, and, simultaneously, generate a new photon with angular frequency $\omega_{TH}$, which satisfies the relationship $\omega_{TH} = 3\omega_{FF}$ due to conservation of energy [6, 41]. The input FF plane wave impinges on the metasurface along the x-y plane with transverse magnetic (TM) polarization and under oblique incident angle $\theta$. The reflected TH radiation is generated by the graphene metasurface and radiated back to free space. In experimental set-ups, the measured radiation power of the generated TH wave will be equivalent to the power outflow across the boundaries of the simulation domain [42-44]. Therefore, the radiated TH power can be described in a more practical way by the power outflow of the TH wave. The conversion efficiency of THG is defined as: $CE = P_{r,TH} / P_{i,FF}$, where $P_{i,FF}$ is the incident power at FF. $P_{r,TH}$ is the radiated TH power outflow calculated by $\int_C \vec{S} \cdot \vec{n}$ with the Poynting vector $\vec{S}$ crossing the boundary curve $C$ multiplied by the boundary norm vector $\vec{n}$ [45]. In all our 2D simulations, we assume that the z-direction length of the graphene ribbons is 1 meter leading to Watt power units.

The properties of bulk materials (dielectric and gold in the proposed metasurface) can be characterized by the relative permittivity $\varepsilon = \varepsilon_L + \varepsilon_{NL}$, where $\varepsilon_L$ and $\varepsilon_{NL}$ are the linear and nonlinear contributions, respectively. In bulk materials, the strength of the third-order nonlinearity is usually measured by the nonlinear susceptibility $\chi^{(3)}$, that relates the nonlinear polarization to the electric field. The induced nonlinear polarization at TH frequency $\omega_{TH}$ is given by [6, 46]:

$$P_{NL} = \varepsilon_0 \chi^{(3)}(\omega_{TH}; \omega_{FF}, \omega_{FF}, \omega_{FF}) E_{FF}^3, \quad (1)$$

where $\varepsilon_0$ is the permittivity of free space and $E_{FF}$ is the electric field induced at the FF mode. In this work, both the gold and dielectric spacer layers are assumed to be isotropic. Hence, the nonlinear relative effective permittivity can be calculated by $\varepsilon_{NL} = P_{NL} / E_{TH}$, where $E_{TH}$ is the electric field of the TH wave. The dielectric material employed in the proposed metasurface has a linear permittivity $\varepsilon_{L,Di} = 4.41$ and Kerr nonlinear susceptibility $\chi_{Di}^{(3)} = 9.96 \times 10^{-20}$ m$^2$/V$^2$, similar to amorphous chalcogenide glass (As$_2$Se$_3$), which is a typical nonlinear dielectric material suitable for far-infrared and low THz nonlinear applications [47]. It will be shown later that graphene has orders of magnitude larger nonlinear susceptibility and is the dominant contribution of the THG achieved by the proposed metasurface. The linear permittivity of gold at THz frequencies is complex and can be obtained using the Drude model: $\varepsilon_{L,Au} = \varepsilon_\infty - f_p^2 / f(f - i\gamma)$, where $f_p = 2069$ THz, $\gamma = 17.65$ THz, $\varepsilon_\infty = 1.53$, and $f = \omega / 2\pi$ is the wave frequency [48]. In the current work, all the computed electromagnetic fields follow the time harmonic convention $[\exp(i\omega t)]$. Note that gold also has a nonlinear susceptibility with value $\chi_{Au}^{(3)} = 2.45 \times 10^{-19}$ m$^2$/V$^2$ at the near-infrared frequency region [43] but the nonlinear properties of gold at low THz frequencies are not available in the scientific literature. This is because gold has a very high conductivity at low THz frequencies and can be approximated as a perfect electric conductor. Hence, the fields do not penetrate inside gold at this frequency regime and cannot interact with its nonlinearity. As a result, the nonlinear properties of gold are irrelevant at low THz and will not affect the total nonlinear response of the metasurface.

Graphene is a 2D conductive material and, as a result, is more accurate to be modeled as a surface current, neglecting its sub-nanometer thickness [49, 50]. It exhibits isotropic properties only along its surface [16]. Due to the THG nonlinear process, the nonlinear surface current of graphene can be expressed by the formula [46]:

$$J = \sigma_L(\omega_{TH}) E_{FF} + \sigma^{(3)}(\omega_{TH}; \omega_{FF}, \omega_{FF}, \omega_{FF}) E_{FF}^3. \quad (2)$$

Thus, the complete material characteristics of graphene can be described by the linear and nonlinear surface conductivities $\sigma_L$ and $\sigma^{(3)}$, which have both complex values. Under the random-phase approximation, the graphene linear conductivity can be derived by the Kubo formula consisting both of interband and intraband electron transitions. However, in this work, we operate at low THz frequencies below the interband transition threshold $\hbar\omega < 2|\mu_c|$, where $\mu_c$ is the Fermi energy or doping level of graphene [46]. In this case, the interband transitions are blocked due to the Pauli principle, so only the intraband transitions exist [38]. The linear conductivity arising from the intraband transitions can be calculated using the Drude dispersion model as: $\sigma_L \approx -iD/\pi(\omega - i/\tau)$, where $D = q^2\mu_c/\hbar^2$, $\tau$ is the electron relaxation time that accounts to the optical loss, $q$ is the electron charge, $\omega$ is the radial frequency, and $\hbar$ is the reduced Planck constant [46, 51]. The linear conductivity of graphene is a function of the wave frequency, electron relaxation time, and the Fermi energy. The electron relaxation time can be affected by many factors, such as temperature, Fermi energy, external field, graphene sample quality, and substrate material [52, 53]. Generally, the electron relaxation time is described by $\tau = \mu\mu_c/qv_F^2$, where $\mu$ is the DC mobility of the graphene sample and $v_F = 1\times 10^6$ m/s is the Fermi velocity [27, 45, 54]. It has been reported that the mobility of graphene can exceed 20 m²/Vs at room temperature [55] and the value of $\tau$ is in the picosecond range [8, 56-59]. Hence, in this work, the electron relaxation time is assumed to be $\tau = 0.5$ ps and we always operate at room temperature. The Fermi energy is initially set to be equal to $\mu_c = 0.3$ eV [56, 57]. The corresponding mobility is relative low in this case and equal to $\mu = 1.67$ m²/Vs.

The third-order nonlinear surface conductivity of graphene at THz frequencies can be calculated by [35]:

$$\sigma^{(3)} = \frac{i\sigma_0(\hbar v_F q)^2}{48\pi(\hbar\omega_{FF})^4}T(\frac{\hbar\omega_{FF}}{2\mu_c}), \qquad (3)$$

where $\sigma_0 = q^2/4\hbar$, $T(x) = 17G(x) - 64G(2x) + 45G(3x)$ with $G(x) = \ln|(1+x)/(1-x)| + i\pi\theta(|x|-1)$, and $\theta(z)$ is the Heaviside step function. The real part of $\sigma^{(3)}$ in Eq. (3) is very small and can be assumed to be negligible at THz frequencies. Hence, only the imaginary part of $\sigma^{(3)}$ determines the strength of graphene's nonlinear performance. In order to have a direct comparison between the nonlinearity of graphene and ordinary bulk materials, the graphene can be assumed to be equivalent to a bulk medium with thickness $d \approx 0.34$ nm and equivalent third-order susceptibility $\chi^{(3)} = -i\sigma^{(3)}/(\omega\varepsilon_0 d)$ [46].

The values of this nonlinear susceptibility are several orders of magnitude larger than conventional dielectric and metallic bulky nonlinear media. Therefore, the large nonlinear response of graphene strongly depends on the fundamental frequency $\omega_{FF} = 2\pi f_{FF}$, where the equivalent third-order susceptibility is approximately proportional to $1/\omega_{FF}^4$. Besides, the nonlinear response is also proportional to $1/\mu_c$. Therefore, a stronger nonlinear graphene response can be expected at lower frequencies and under low Fermi energy or doping level. The values of the equivalent nonlinear susceptibility of graphene are very large $\chi^{(3)}_{graphene} = 1.3 \times 10^{-10}$ m$^2$/V$^2$ at 3 THz, while the nonlinear susceptibility of gold at near-IR is $\chi^{(3)}_{Au} = 2.45 \times 10^{-19}$ m$^2$/V$^2$ [39] and the nonlinear susceptibility of the dielectric material is on the order of 10$^{-18}$-10$^{-20}$ m$^2$/V$^2$, both used in the proposed nonlinear metasurface design. Therefore, the nonlinear response of the metasurface will be dominated only by the ultrastrong Kerr nonlinear susceptibility of graphene, as it will be demonstrated later by using full-wave simulations. Furthermore, the fields are mainly confined along the graphene surface and are much lower inside the dielectric spacer layer. The high nonlinear properties of graphene makes it a very appealing nonlinear material in THz frequencies.

Beside the values of nonlinear susceptibility or conductivity, the generated TH power is also affected by the field intensity at the FF mode. The FF mode is locally confined and dramatically enhanced at the resonance by exciting SPP standing waves along the graphene micro-ribbons. The incident FF wave must be TM polarized in order the proposed metasurface to resonate, where the magnetic field direction is along z-axis [23, 54, 60, 61]. Moreover, the fields at the FF mode are further enhanced inside the dielectric spacer layer and along the graphene micro-ribbons by introducing the metallic substrate termination. This large field enhancement is greatly favorable in order to boost the THG efficiency. Furthermore, THG, as a typical parametric process, also depends on the phase mismatch between FF incident and TH generated waves. However, the phase-matching requirement is greatly relaxed along the highly subwavelength area of the proposed graphene metasurfaces [30-33]. Finally, it is noteworthy to mention that the current work is focused only on THG and all the other relevant nonlinear processes, such as higher harmonic generation (HHG), are ignored because their generated frequencies will be separated from the TH frequency.

## 3. Results and discussion

To investigate the resonant response and field enhancement caused by the proposed graphene metasurface, we used the dimensions and parameters given in the previous section ($a = 3.88$ μm and $d = 0.75a$) and varied the thickness of the dielectric layer $t$. The reflectance simulation results are shown in figure 2(a), where the resonant frequency of the proposed metasurface is blue-shifted with the decrease of the dielectric thickness. The field enhancement induced along the metasurface by the incident FF wave is calculated by the maximum ratio $|E_{FF}/E_0|_{max}$, where $E_0$ represents the incident electric field amplitude of the input FF wave.

We calculate and plot in figure 2(b) the maximum field enhancement for dielectric layers with different thicknesses. The field is boosted at the resonance frequency due to the excitation of SPP along the graphene metasurface. Note that there is a minor resonant frequency shift between the peak of the field enhancement [figure 2(b)] and the minimum reflectance [figure 2(a)]. This is attributed to the small detuning introduced by the thin dielectric spacer layer combined with the highly confined surface plasmons along the graphene surface, which was also previously obtained with materials with strong plasmonic response based on noble metals [62, 63]. The same phenomenon was also observed before in [45] and was attributed to the phase shift induced by the nonlinear Kerr effect. However, our current work proves that the small frequency shift originates from the linear characteristics of the graphene micro-ribbon metasurface. When $t = 6.3\,\mu m$, perfect absorption will occur at the resonant frequency of 3 THz, where the reflectance is zero $(R = 0)$ and the incident wave is perfectly absorbed by the metasurface $(A = 1 - R = 1)$. Such a large absorption is unexpected, considering the extreme subwavelength thickness $(t \simeq \lambda/25)$ of the proposed structure. The peak field enhancement can reach a maximum value at this resonant condition and the electric field of the FF mode can be enhanced by nearly 160 times [see figure 2(b)]. The peak of the electric field always coincides with the resonance frequencies for the same graphene properties and different thickness metasurfaces, as it can be seen in figures 2(a) and 2(b). It is important to mention that larger electric field values will lead to stronger THG according to the relationship given by Eq. (2).

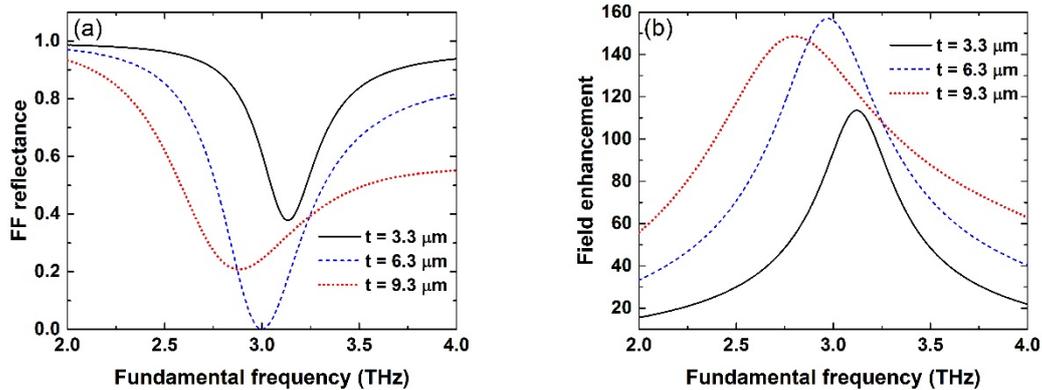

Figure 2. Computed (a) reflectance and (b) field enhancement of the FF mode at linear metasurface operation as functions of the incident frequency for various thicknesses of the dielectric spacer layer.

To utilize the strong field enhancement in the THG process, the frequency of the FF excitation wave should be located as close as possible to the resonance. For a spacer layer with optimized thickness ($t = 6.3\,\mu m$), the resonant frequency is computed to be located at 3 THz, where the reflectance is minimum and the absorption is maximum. The FF input wave illuminates the proposed metasurface at normal incidence with relative low input intensity

$I_{i,FF} = 30$ kW/cm$^2$. This value is much lower compared to the GW/cm$^2$ radiation used to illuminate graphene at previous experimental works [16, 18] and is not expected to thermally affect and degrade the presented graphene structure. Figure 3 demonstrates the power outflow of the TH wave, which is going to be at 9 THz, as a function of the incident wave's frequency. When the FF input wave approaches the resonant frequency (3 THz), the TH radiation power significantly increases with a maximum value of 0.26 W. The TH radiation curve shown in figure 3 follows the same trend with the FF field enhancement illustrated in figure 2(b). Thus, the peak of the maximum TH radiation power is slightly shifted (with a shift computed to be 25 GHz) compared to the main resonance (3 THz) due to the detuning reasons described before. The insets in figure 3 demonstrate the calculated normalized electric field distributions along two neighboring graphene micro- ribbons at the FF mode equal to 3 THz (upper inset) and the corresponding TH mode at 9 THz (lower inset). Interestingly, a strong localized plasmon resonance is formed at the FF mode and the electric field is dramatically enhanced on the edges of the graphene ribbons. On the contrary, the TH field distribution is off resonance, which facilitates the efficient out-coupling of the nonlinear TH radiation back to free space.

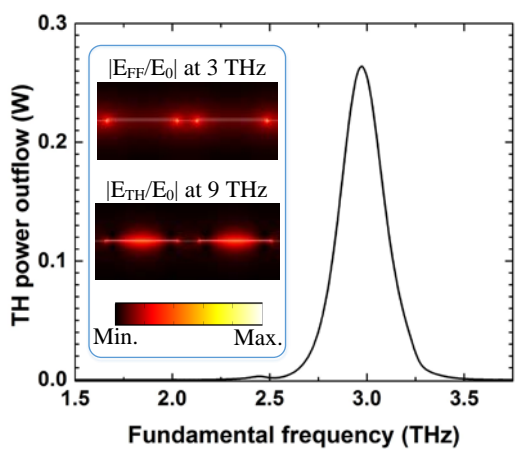

Figure 3. Computed power outflow of the TH radiation as a function of the incident wave frequency. Inset: The computed normalized electric field distributions around the graphene micro-ribbons at the FF mode equal to 3 THz (up) and the corresponding TH frequency at 9 THz (down).

Next, the dependence of the TH radiation on the incident angle is further analyzed. The FF input wave is launched at 3 THz with a varying incident angle between -88° to +88° and the results are shown in figure 4. Higher reflectance is obtained as the incident angles are increased. The computed reflectance is demonstrated in figure 4 (black line), indicating that the coupling of the proposed metasurface to free space radiation is weaker, i.e. the energy coupled to the metasurface is reduced at oblique incidence angles. This leads to reduced TH power outflow at oblique incidence and the computed THG result is plotted in figure 4 (red line). However, the TH radiation can maintain larger than half of its peak values when the incident angles are in the range of -36° to +36°.

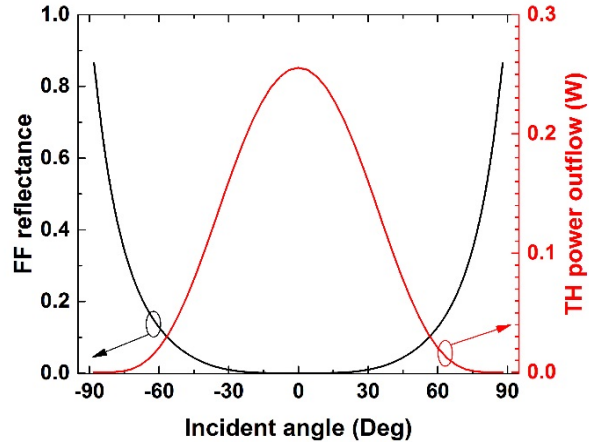

Figure 4. Effect of incident angle on the FF wave reflectance (black line) and the TH power outflow (red line), where the FF incident wave is excited at 3 THz.

The TH radiation is computed under different material scenarios and the results are compared in figure 5. The TH power outflow of the proposed nonlinear graphene metasurface (black solid line) has the largest value in figure 5 compared to all the other scenarios: a) same dimensions graphene micro-ribbon metasurface on top of the same thickness dielectric layer but without gold substrate (purple dash-dotted line), b) a uniform graphene monolayer film placed on top of the dielectric layer without gold substrate (red dotted line), and c) a bare gold film with a dielectric top layer and without graphene (blue dashed line). The thickness of the dielectric layer is always kept constant in all the aforementioned cases for fair comparison. The structure composed of graphene micro-ribbons on top of a dielectric layer but without the gold substrate termination (figure 5, purple dash-dotted line) is similar to the metasurface proposed in a recent work [45]. Under normal incidence, the TH radiation generated by the proposed graphene metasurface is *thirteen orders of magnitude larger* compared to the bare gold film with a dielectric top layer and no graphene, five orders of magnitude larger than the uniform graphene case, and two orders of magnitude larger compared to the same graphene micro-ribbons metasurface operating in transmission when the metallic substrate is removed, similar to the structure proposed in [45]. Note that all these results are obtained for the same 3 THz incident wave to achieve a fair comparison. The currently proposed graphene metasurface has the highest reported THG efficiency compared to all relevant existing works [8, 9, 16, 18, 39, 45]. Finally, we replaced the graphene ribbons with similar gold stripes and we computed the THG performance which was found to be very weak (not shown here) and comparable to the bare gold film with a dielectric top layer performance (blue dashed line in figure 5). This is expected because gold's nonlinear susceptibility is orders of magnitude smaller compared to graphene's nonlinear susceptibility [39].

To further prove that the gold's nonlinear properties will not affect the THG performance of the graphene metasurface, we swept the nonlinear susceptibility of gold $\chi_{Au}^{(3)}$ to a hypothetical range of values from very high $10^{-16}$ m$^2$/V$^2$ to very low $10^{-22}$ m$^2$/V$^2$ and performed identical simulations with the results presented in figures 3 and 4. It turns out that the TH power outflow of the proposed graphene metasurface remains unchanged independent of the nonlinear susceptibility of gold (results not shown here). As it was mentioned before, the fields do not penetrate and interact with gold at low THz frequencies, leading to an extremely small optical interaction volume and negligible THG. This is the main reason why gold is not practical for low THz nonlinear applications. In the proposed metasurface design, it is just used as a perfect metallic substrate to achieve zero transmission and operation in reflection.

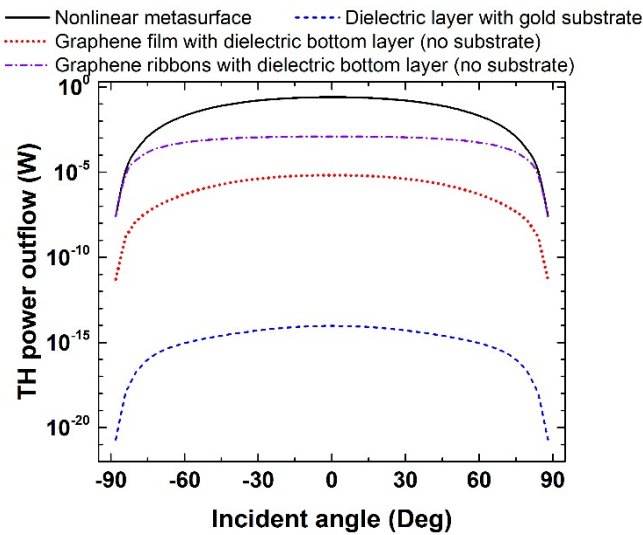

Figure 5. Comparison of the TH power outflow for different incident angles among the proposed graphene metasurface (black solid line), a bare gold film with a nonlinear dielectric top layer and without graphene (blue dashed line), a uniform graphene monolayer film with a nonlinear dielectric bottom layer without gold substrate (red dotted line), and the same graphene micro-ribbon metasurface operating in transmission with a nonlinear dielectric bottom layer and without gold substrate (purple dash-dotted line). All the structures are excited with 3 THz FF incident waves.

The generated TH power outflow should be proportional to the third power of the incident FF power in the THG nonlinear process. This interesting property provides a convenient way to adjust, tune, and control the output power of the generated TH radiation. We verify this relationship based on simulations with results plotted in figure 6(a), where the FF wave (3 THz) is normal incident to the proposed metasurface. It is demonstrated that the computed TH power outflow can be well fitted by using a cubic power function $aI_{i,FF}^3$ with respect to the incident FF mode intensity $I_{i,FF}$, where $a = 9.5 \times 10^{-27}$ m$^5$/W$^2$ is the calculated fitting parameter derived from the simulation results shown in figure 6(a). In addition to the TH radiation, the THG power conversion efficiency is also dramatically improved by increasing the incident FF wave intensity, as it is shown in figure 6(b). Interestingly, the THG conversion efficiency can reach high values (-26 dB) when relative low FF input intensities ($I_{i,FF} = 0.1$ MW/cm$^2$) are used to excite the proposed nonlinear graphene metasurface.

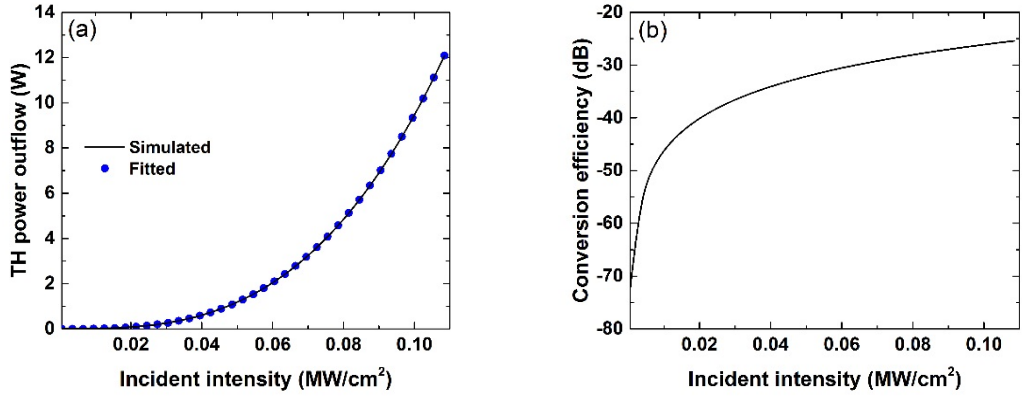

Figure 6. Computed (a) TH power outflow and (b) THG conversion efficiency as functions of the incident FF wave intensity. The TH power outflow is fitted by using a cubic power function presented by the blue dotted line.

The conductivity of graphene can be made tunable as the Fermi energy $\mu_c$ is adjusted by the applied voltage between graphene and metallic substrate. Note that in this configuration the relationship between $\mu_c$ and the bias voltage $V_g$ can be described as $\mu_c = h v_F \sqrt{\pi C V_g / q}$, where $h$ is the Planck constant, and $C = \varepsilon_{L,Di} \varepsilon_0 / t$ is the electrostatic capacitance per unit area [64]. In our case, the maximum bias voltage is computed to be 81 V for a Fermi level of 0.4 eV and $t = 6.3$ μm dielectric layer thickness. This voltage value is realistic for a potential experimental verification of the proposed structure and below the dielectric breakdown voltage [25, 65].

As it was previously discussed, the mobility $\mu$ determines the electron relaxation time $\tau$ of graphene properties under different Fermi level $\mu_c$ values. However, different graphene samples may have dissimilar values of mobility due to different fabrication procedures. The effect of the graphene mobility on the linear and THG process is demonstrated in figure 7, where the Fermi energy is kept constant at $\mu_c = 0.3$ eV and the FF wave is perpendicularly incident to the proposed metasurface. The electron relaxation time changes from 50 fs to 5 ps, which corresponds to mobility values from 0.167 m²/Vs to 16.7 m²/Vs. The variation of $\tau$ affects the real and imaginary parts of the graphene conductivity in such a way that the resonant frequency of the metasurface is kept almost constant, as it is presented in figure 7(a). Nevertheless, by increasing $\tau$, the reflectance at the FF resonance first decreases and then increases, suggesting that the coupling to the metasurface is increased and then decreased, respectively. On the other hand, the field enhancement peak of the FF wave is monotonously increasing with respect to $\tau$ at the resonant frequency and the result is depicted in figure 7(b). This is expected because the graphene has lower losses for larger relaxation times. The raise of the field enhancement at the same resonating frequency can lead to more intensified, enhanced by 34-fold, TH radiation depicted in figure 7(c).

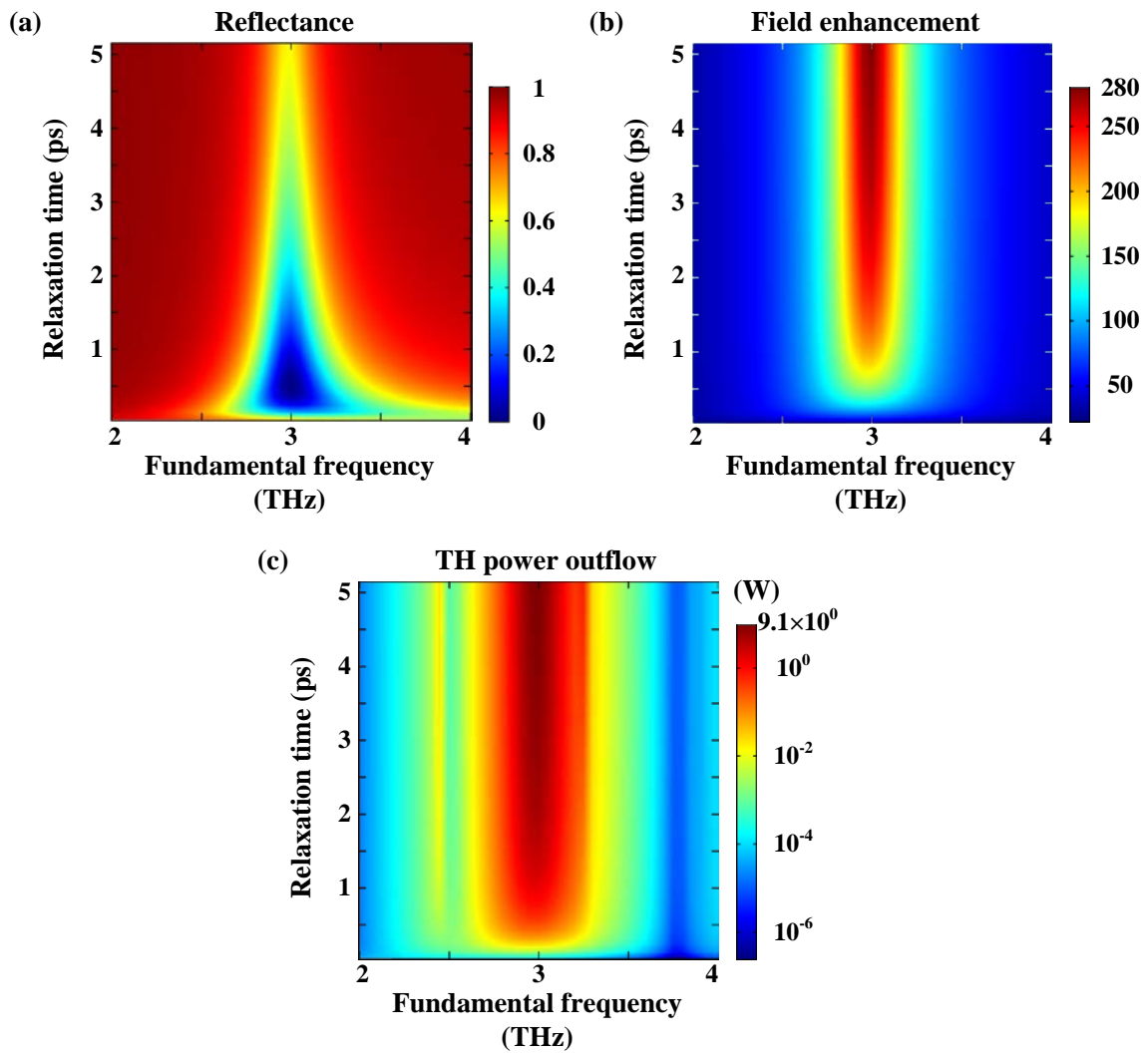

Figure 7. Computed distributions of (a) FF linear reflectance, (b) FF linear field enhancement, and (c) TH power outflow as functions of the fundamental frequency and graphene's electron relaxation time (i.e. mobility).

In figure 8, we plot the distributions of FF reflectance [8(a)], FF field enhancement [8(b)], and TH power outflow [8(c)] when $\mu_c$ varies from 0.05 eV to 0.4 eV and the FF incident wave frequency is swept from 1.5 THz to 3.5 THz. At this frequency range, the condition $\hbar\omega < 2|\mu_c|$ is always satisfied for both FF and TH waves, which guarantees the absence of interband transitions and the use of graphene conductivity formulas presented before. Moreover, we always consider normal FF incident wave angle $(\theta = 0°)$ and the graphene's mobility is kept constant to a moderate value $\mu = 1.67$ m²/Vs to calculate the results presented in figure 8. The Fermi energy has large impact on the reflectance spectrum of the FF mode, as it is demonstrated in figure 8(a). The variation of the Fermi energy not only alters the reflectance at the resonance, but also considerably shifts the resonance frequency of the proposed metasurface. In particular, the resonance is blue-shifted as $\mu_c$ is increased, while the reflectance values at resonance are decreased and then increased out of resonance. The reflectance is equal to zero for $\mu_c = 0.3$ eV at the resonant frequency ($f_{FF} = 3$ THz), where critical coupling occurs. The field enhancement of the FF mode is illustrated in figure 8(b) as a function of the fundamental frequency and the Fermi energy. When the Fermi energy is increased, the peak value of the field enhancement located close to the resonant frequency is also enlarged.

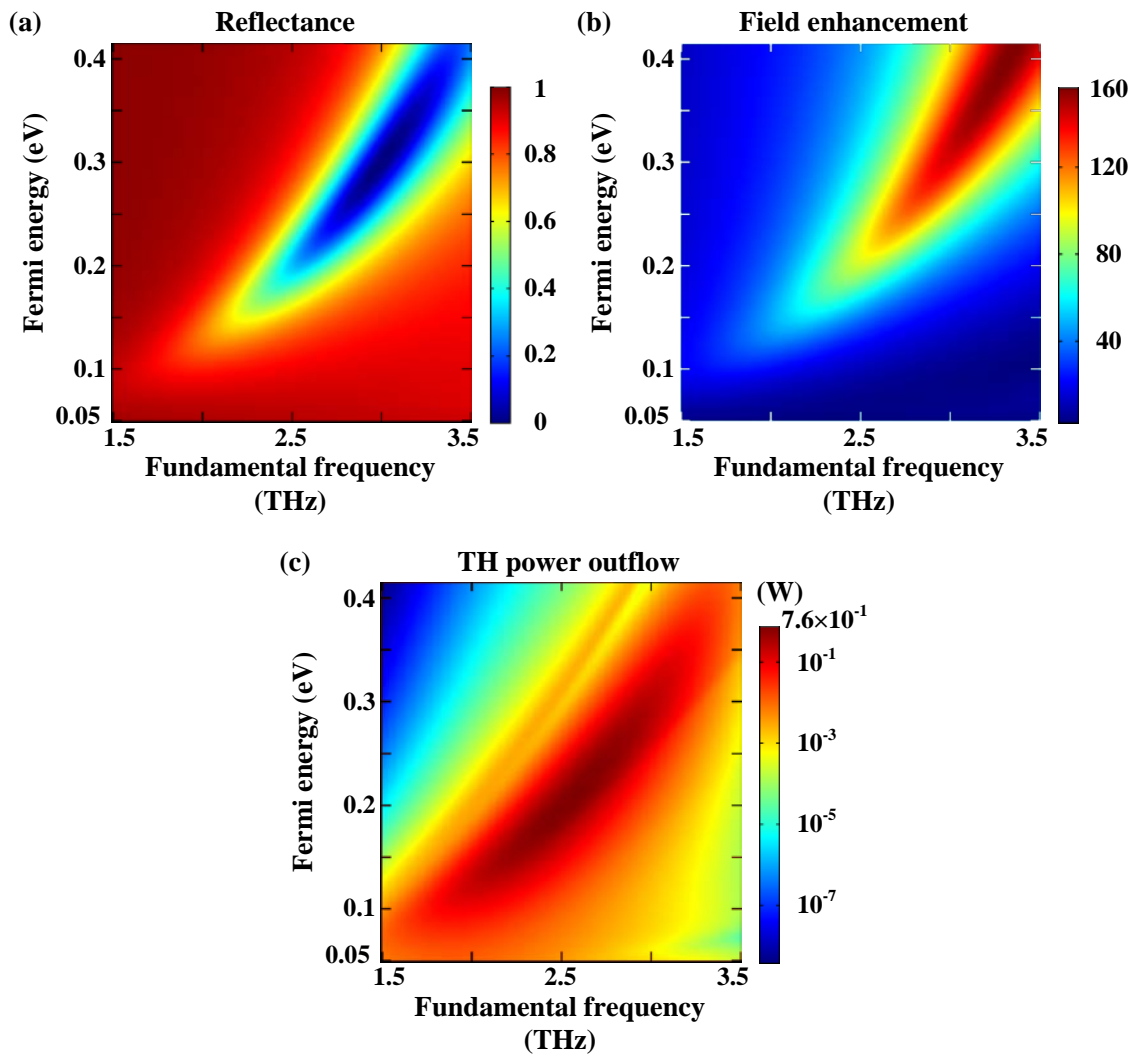

Figure 8. Computed distributions of (a) FF linear reflectance, (b) FF linear field enhancement, and (c) TH power outflow as functions of the fundamental frequency and graphene's Fermi energy.

Figure 8(c) represents the effect of the Fermi energy on the TH radiation spectrum. When adjusting $\mu_c$, the frequency corresponding to the TH radiation peak approximately follows the variation of the resonance depicted in figure 8(a). However, the maximum TH radiation while sweeping $\mu_c$ and $f_{FF}$ does not take place at the critical coupling frequency point, as it can be seen by comparing the figures 8(a) and 8(c). This interesting result is due to the high dispersive nature of graphene's nonlinear conductivity combined with the excitation of higher-order resonances close to the TH frequency of the nonlinear metasurface. Equation (3) demonstrates that the increase of both Fermi energy and resonant frequency will lead to a large reduction in the values of graphene's nonlinear conductivity and, as a result, to lower TH radiation. In addition, higher-order resonances will improve the out-coupling of the TH radiation if they are excited close to the THG frequency. Due to the reasons, it is found [figure 8(c)] through accurate full-wave nonlinear simulations that the maximum TH radiation will be stronger at slightly lower frequencies and Fermi energy values compared to the critical coupling frequency point. Note that when the coupling with the nonlinear metasurface is totally lost [figure 8(a), below 1.75THz and 0.1eV], then the TH radiation power also abruptly drops, as it is obtained in figure 8(c). This interesting result will help in the future experimental verification of the proposed nonlinear graphene metasurfaces. The maximum TH radiation power outflow is 0.76 W and occurs at $f_{FF} = 2.55$ THz with $\mu_c = 0.2$ eV [figure 8(c)].

Hence, by just tuning graphene's Fermi energy, the proposed tunable nonlinear metasurface can be optimized to achieve maximum TH radiation at different frequencies of the incident FF wave [figure 8(c)] without altering its geometry and dimensions. It is expected that the tunability of the proposed nonlinear graphene metasurface will offer great flexibility and new adaptable functionalities in future THz applications that cannot be achieved with the usual nonlinear plasmonic metallic metasurfaces.

Finally, we would like to stress that in the presented simulation results, we included both the linear and nonlinear properties of all the materials involved, such as graphene, dielectric spacer layer, and gold substrate. This made our simulations more realistic and closer to the results expected by a future experimental verification of the proposed concept. To further investigate which material will dominate the THG process of the proposed nonlinear graphene metasurface, we calculate in figure 9 the TH radiation while separately ignore the nonlinearity of each different material of the nonlinear structure. All the other parameters are kept the same to the ones used before in the results presented in figure 4. When we remove the nonlinear properties of the dielectric layer and gold substrate ($\chi_{Di}^{(3)} = \chi_{Au}^{(3)} = 0$), the decrease in the TH radiation is too small to be identified compared to the case when the nonlinear properties of all materials are included. On the contrary, if we remove the nonlinear properties of graphene by setting its third-order conductivity equal to zero ($\sigma^{(3)} = 0$), the computed TH power outflow is decreased by at least ten orders of magnitude. Therefore, we can conclude that the source of the nonlinear radiation is mainly the patterned graphene micro-ribbons. The nonlinear effects of the dielectric layer and gold substrate are minor at THz frequencies. One straightforward reason to explain this effect is that the nonlinear conductivity of graphene is considerably larger in the THz frequency range compared to the other materials. In addition, the incident FF wave is enhanced and confined only around the graphene micro-ribbons (see inset of figure 3) due to the induced surface plasmons at the resonance frequency. Hence, the proposed new nonlinear graphene metasurface design can enhance THG at THz frequencies with efficiencies that cannot be achieved by any other material with the same subwavelength dimensions.

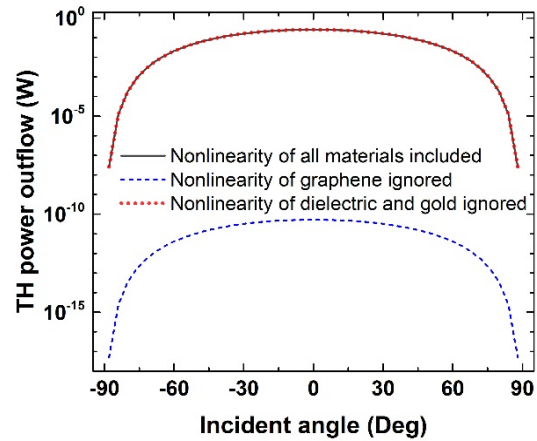

Figure 9. Computed TH power outflow of the nonlinear metasurface when the nonlinear properties of all materials are considered (black solid line), only the nonlinearity of graphene is ignored (blue dashed line) or the nonlinearities of gold and dielectric are ignored (red dotted line).

## 4. Conclusions

Suffering from poor efficiency, enormous input intensities are usually required to trigger THG from subwavelength structures. In this work, a new nonlinear ultrathin metasurface is proposed based on periodically patterned graphene micro-ribbons to obtain THG at THz frequencies with low input intensities and high efficiencies. The electric field is strongly enhanced at the resonance of the metasurface due to the excitation of surface plasmon and FP resonances along its surface and inside the dielectric substrate, respectively. In addition, graphene has much stronger third-order nonlinear conductivity at far-infrared and THz frequencies compared to the vast majority of dielectrics and noble metals. Hence, when the incident wave is launched at the resonance frequency of the proposed metasurface, the THG conversion efficiency is dramatically enhanced by several orders of magnitude. The maximum conversion efficiency occurs when the FF excitation wave impinges at normal incident angle. However, the incident angle can be varied in a considerably broad range without greatly decreasing the TH radiation outflow leading to a relative angle-independent response. We had also verified, based on full-wave simulations, the cubic power relationship between the TH radiation power outflow and the incident intensity used to excite the ultrathin metasurface. Interestingly, the conversion efficiency can reach unprecedented high levels of approximate -26 dB at low THz frequencies when the incident intensity has relative low values on the order of 0.1 MW/cm$^2$. The need of low intensity values to obtain strong THG at THz frequencies is a major advantage of the proposed graphene metasurface. It is expected that it will make graphene THG at low THz frequencies more practical, considering that high input power THz sources do not exist [5].

In addition, the interesting reconfigurable and tunable properties of the proposed nonlinear graphene metasurface are analyzed in detail. It is found that the Fermi energy or doping level of graphene can shift the resonance frequency, nonlinear conductivity, and optical losses of the proposed metasurface. By dynamically tuning the Fermi energy, the tunable nonlinear graphene metasurface can emit TH radiation at different THz frequencies without changing its geometry and dimensions. Finally, we also demonstrated that the patterned graphene configuration is the main source of the nonlinear radiation. The proposed ultrathin tunable graphene metasurface designs provide a new method to introduce extremely large nonlinearity enhancements at THz frequencies, allowing a huge reduction in the required input intensities. In particular, they can enhance THG with efficiencies that cannot be achieved by any other material with the same subwavelength dimensions. The reported results pave the way to the design of future nonlinear chip-scale THz devices that can be used as tunable and adaptive THz and infrared sources or wave mixers.

## Acknowledgments

This work was partially supported by the Office of Research and Economic Development at University of Nebraska-Lincoln, the National Science Foundation (NSF) through the Nebraska Materials Research Science and Engineering Center (MRSEC) (grant No. DMR-1420645), and the Nebraska's Experimental Program to Stimulate Competitive Research (EPSCoR).